\documentstyle[twocolumn,aps,epsf]{revtex}

\begin{document}
\title{Labyrinthic Granular Landscapes}

\author{H. Caps and N.Vandewalle}

\address{GRASP, Institut de Physique B5, Universit\'e de Li\`ege, \\
B-4000 Li\`ege, Belgium.}

\address{~\parbox{14cm}{\rm\medskip We have numerically studied a model of granular landscape eroded by wind. We show the appearance of labyrinthic patterns when the wind orientation turns by $90^\circ$. The occurence of such structures are discussed.  Morever, we introduce the density $n_k$ of ``defects'' as the dynamic parameter governing the landscape evolution. A power law  behavior of $n_k$ is found as a function of time. In the case of wind variations, the exponent (drastically) shifts from 2 to 1. The presence of two asymptotic values of $n_k$ implies the irreversibility of the labyrinthic formation process.  \\ \date{\today}}}

\maketitle

\hspace{-13pt} 45.70.Mg, 81.05.Rm, 64.75.+g


\section{Introduction}

The ripple formation due to the wind blowing across a sand bed \cite{bagnold} has recently received much attention in the statistical physics community \cite{hoyle,werner}. Indeed, the physical mechanisms involved are complex phenomena of granular transport. Experi ental works as well as natural observations \cite{bagnold} have underlined the primary role played by saltation in the emergence of ripples and dunes. Along this line, various models for ripple formation have been proposed in the past. Theoretical models which consider hopping and rolling grains have generally led to traveling ripple structures \cite{hoyle}. Simulations \cite{werner,nishimori,ijmpc} have also considered various additional effects like the screening of crests, grain reptation and the existence of a grain ejection threshold. Among others, the Nishimori-Ouchi (NO) model \cite{nishimori} is able to reproduce a wide variety of different eolian structures: transverse, barkhanic, star-like, etc... The main advantage of such numerical model is that all parameters can be easily tuned and the physical mechanisms can be deeply investigated.

In 1991, Goossens \cite{goossens} performed an original experiment which is the following. In a wind tunnel, a $12 \times 12$ $cm^2$ rough granular landscape (dust size $\approx 32\, \mu\, m$) is eroded by an air flow  ($v \approx 132\, cm \, s^{-1}$). This created small ripples perpendicular to the wind direction. After 45 minutes, the wind orientation was changed by $90^{\circ}$ and this poduced diagonal structures instead of a new set of ripples perpendicular to the previous ones. The Goossens's experiment represents a good test for the NO model. In the present paper, we report simulations of this particular kind of landscape. This allows us to discuss the dynamics of this unusual phenomenon.

\section{Model}

In the Nishimori-Ouchi model \cite{nishimori}, two kinds of granular transport
processes are considered: {\it (i)} the saltation and, {\it (ii)}
the potential energy relaxation. The temporal evolution equation of
the height of sand $h(x)$ at point of coordinate $x$ reads
\begin{equation}\label{no}
\frac{\partial h(x,t)}{\partial
t}=A\left(N(\ell)\frac{d\ell}{dx}-N(x)\right)+D\frac{\partial^2h(x,t)}{\partial
x^2}.
\end{equation}
At the right hand side of this equation, the first term represents the saltation process.
Due to wind shear stress, grains are moved from a position
$\ell(x)$ to a position $x$. The mean amplitude of the path length
is given by the constant $A$. On the other side, the constant $D$
is a relaxation coefficient. This second term takes in account the
transport phenomena along the slopes of the surface, e.g. reptation
and avalanches.

The NO model has been implemented as follows. A two
dimensional square lattice with periodic boundary conditions is
considered. To each site $i,j$ of the lattice is associated a
real number $h_{i,j}$ which represents the height of the granular
landscape at that position. Assume that the wind blows along the
$i$-axis. At each discrete time $t$, a site $i,j$ is randomly
choosen and a quantity $q_{i,j}$ of matter is displaced by
saltation from this site towards the site $i+\ell_{i,j},j$ which
is incremented by the height $q_{i,j}$. Both quantities
$\ell_{i,j}$ and $q_{i,j}$ are determined by
\begin{eqnarray}\label{salt}
\ell_{i,j} & = & \alpha (\tanh \nabla h_{i,j} + 1) \cr q_{i,j} &
= & \beta (1 + \epsilon - \tanh \nabla h_{i,j})
\end{eqnarray} where $\alpha$ and $\beta$ are dynamical constants and the
parameter $\epsilon$ is the minimum quantity of sand which is
displaced by saltation. The mathematical form ($\tanh \nabla h$) of those
relationships assumes that the local slope mainly controls the
granular transport. The flux of sand extracted from the faces
exposed to the wind is indeed smaller than that screened by
crests. After the saltation process (\ref{salt}), a relaxation of the
landscape is assumed (creeping and avalanches) before the next
time step $t+1$ takes place. The relaxation reads


\begin{eqnarray} 
&h_{i,j}(t+1)=h_{i,j}(t)\nonumber\\&+D \left( \frac{1}{6} \sum_{nn}h_{nn}(t) + \frac{1}{12} \sum_{nnn} h_{nnn}(t) - h_{i,j}(t)\right)\nonumber  
\end{eqnarray}

\begin{eqnarray} 
&h_{i+\ell_{i,j},j}(t+1)=h_{i+\ell_{i,j},j}(t)
\nonumber\\&+D \left( \frac{1}{6} \sum_{nn} h_{nn}(t) + \frac{1}{12}
\sum_{nnn} h_{nnn}(t) - h_{i+\ell_{i,j},j}(t) \right)\nonumber
\end{eqnarray} where the summations run
over nearest neighbors ($nn$) and next nearest neighbors ($nnn$)
of both sites $i,j$ and $i+\ell_{i,j},j$. This equation is the
discrete counterpart of the Laplacian relaxation of Eq.(\ref{no}).
The process is repeated a large number of times. Typically, we
stop the simulation after $t=2.5\times 10^{7}$ steps on $201 \times 201$
lattices. We intentionaly choose a lattice size that is not commensurable 
with the mean saltation lenght, $\alpha$. 

\section{Results and discussion}

We have performed extensive simulations by varying all the parameters: $\alpha$, $\beta$, $D$ and $\epsilon$. Modifying $\alpha$ changes the mean ripple wavelength (the distance between two sucessive crests), while $D$ affects the aspect ratio (amplitude) of the ripples. The values taken by $\beta$ and $\epsilon$ permit or not the appearence of ripples: typicaly $\beta\in[0.2,0.6]$ and $\epsilon=0.3$. Note that we have normalized the time $t$ by the duration of the simulation, involving $t\in[0,1]$ in arbitrary units (a.u.).  

\begin{figure}[h]
\centerline{\epsfxsize=7.cm
\epsffile{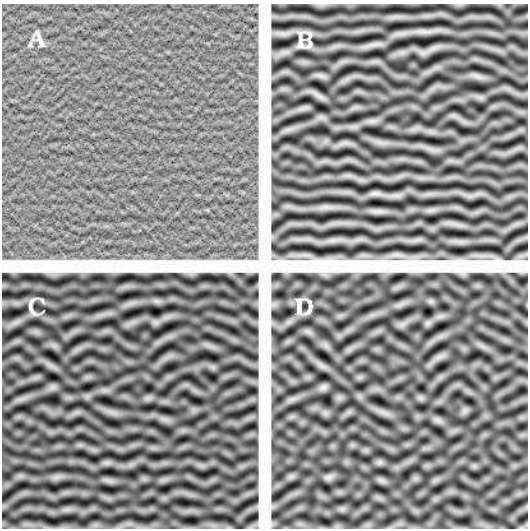}}
\vskip 0.2true cm
\caption{Four different stages of a granular landscape evolution within the NO model. When the wind direction changes, a labyrinthic pattern appears. The simulation parameters are: $\alpha=$2.5, $\beta$=5, $D$=0.4, $\epsilon=0.3$. The lattice size is 101$\times$101. {\it (a)} t=0.2 a.u. and wind direction is up, {\it (b)} t=0.48 a.u. and wind direction is up, {\it (c)} t=0.52 a.u. and wind direction is right, {\it (d)} t=1 a.u. and wind direction is right.}\label{fig1}
\end{figure}

Figure \ref{fig1} shows a typical result of our simulations for $\alpha=2.5$, $\beta=5$, $D=0.4$, $\epsilon=0.3$. The granular landscape is shown for 4 different stages of evolution. One observes on the top row the early formation of ripples perpendicular  to the wind direction. On the bottom row, the wind direction has changed by $90^\circ$ clockwise and a labyrinthic structure appears. This observation emphasizes the impact of the initial topography on the orientation of the ripples crests. One should note that such labyrinthic structures are strikingly similar to Goossens' ones \cite{goossens}.

In order to quantify the effect(s) of wind variations, we have measured the maximum ripple amplitude $A_{max}$ for both constant and variable winds. This quantity was also experimentally measured in \cite{goossens}. As the wind orientation is changed by 90$^\circ$, no significant change of $A_{max}$ is observed, similarily to experiments \cite{goossens}. This means that a brutal change in the wind direction does not modify the net deposit of sediments on the crests. The competition between transport and deposit phenomena is not deteriorated by the perturbation of the wind orientation. Actually, $A_{max}$ represents adequately this competition, but is not a relevant dynamical parameter in order to understand the formation of labyrinthic structures. 

Looking for details in Figure \ref{fig1}, one can observe that: {\it (i)} diagonal structures appear at the vicinity of ``defects" of the primary landscape. Those ``defects'' are {\it kinks} and {\it antikinks}. A kink is a bifurcation of a crest, while an antikink is a termination of a crest, i.e. a bifurcation of the valleys. Moreover, kinks and antikinks are not independent at all: they are formed by pairs. Kinks and antikinks can be considered as nucleation centers for new ripples when the wind direction changes. One understands the formation of labyrinths as follows. Old ripples are pushed in the new wind direction. If their crests are perpendicular to the new direction ripples are compressed. However, near a ``defect'', the angle between the crest and the wind is smaller than $90^\circ$. A rotation of the crest is thus initialized there. This leads to diagonal structures. {\it (ii)} The formation of a labyrinthic-like structure involves the growth of the number of ``defects''.

\begin{figure}[h]
\centerline{\epsfxsize=7.cm
\epsffile{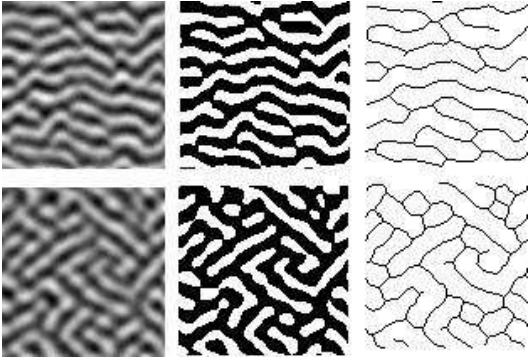}}
\vskip 0.2true cm
\caption{Illustration of the procedure of kink counting in the case of constant wind {\it (top row)} and variable wind {\it (bottom row)}. Grayscale images are created with crests in black {\it (left)}. The binary (black/white) images are extracted and show the crests {\it (midle)}. An iterative erosion technique leads to a skeleton {\it (right)}.}\label{fig3}
\end{figure}

Let us consider the relevant parameter: the density $n_k$ of kinks. This quantity is defined as the number of kinks present on the surface, divided by the area of the lattice. In order to measure the number of kinks present in the landscape, we proceeded as follows (see the illustration in Figure \ref{fig3}). The surface is recorded in grayscale images at different stages of evolution. The darkness indicates the height of sand, i.e. crests are in black while valleys are in white. Images are then analysed using common tools of image analysis. First, a threshold is applied in order to get binary (black/white) images with crests in black. Then, a function reduces all crests to a skeleton through an iterative erosion technique. The last step concerns the countdown itself. The program browses the skeleton line by line. When a black point is met (a crest), the number of its black neighbors is counted. If this number is greater or equals to 3, the point is necessary a kink. One should note that this method can be applied to images of real experiments.

\begin{figure}[h]
\centerline{\epsfxsize=9.cm
\epsffile{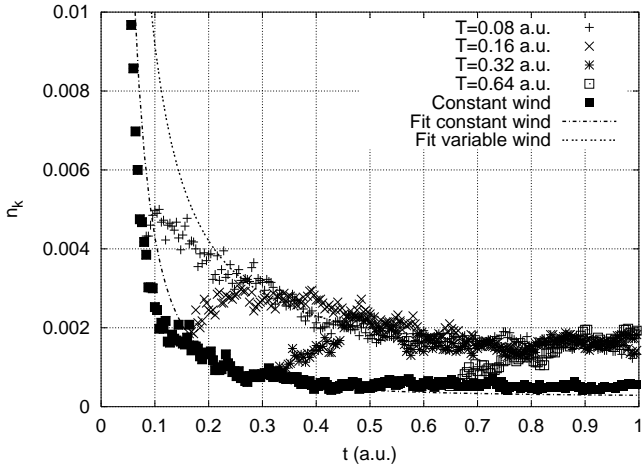}}
\vskip 0.2true cm
\caption{The density $n_k$ of kinks as a function of time $t$. The simulation parameters are : $\alpha=$2.5, $\beta$=5, $D$=0.4, $\epsilon=0.3$, the lattice size is 201$\times$201. Different times $T$ for wind orientation changes are illustrated. The curves are fits for both cases: constant wind (bottom curve) and variable wind (top curve).}\label{fig5}
\end{figure}

Figure \ref{fig5} presents the temporal evolution of $n_k$ in both cases: constant and variable wind. One can see that without any wind modification, $n_k$ decreases as a function of time (black squares). When a wind change occurs at time $T$, the density of kinks suddenly increases. In Figure \ref{fig5} different wind variations are illustrated for different times $T$. After the jump of $n_k$ observed at time $T$, the kink density continues to decreases slowly.

Since the kink density decreases in a faster way than a logarithmic-like law, and in a slower way than an exponential one, we have assumed a power law decay 
\begin{equation}\label{fiteq}  
n_k(t)=a+\frac{b}{c+t^d}
\end{equation}

where $a$, $b$, $c$ and $d$ are fitting parameters. The parameter $a$ represents the asymptotic density of kinks, while the exponent $d$ captures the dynamics of the landscape. Indeed, a large value of $d$ involves a fast decrease of the kink density; such a situation implies that the surface can be easily modified by the wind. On the other hand, a small value of $d$ means that the patterns are less affected by the variation of the wind orientation. Looking for details in Figure \ref{fig5}, one should note that the wind variation does not affect the decay law of $n_k$ after the jump. Actually, the density of kinks follows Eq.(\ref{fiteq}) before and after the wind change.

Typical fits using Eq. \ref{fiteq} are drawn in Figure \ref{fig5} and parameters are reported in Table 1. 
The upper line of Figure \ref{fig5} correspond to the case of a variable wind, and is caracterized by $d\approx 1$, while the lower curve is for a constant wind and follows $d\approx 2$. This difference implies a greater stability of the labyrinthic structure, and the existence of two modes. 

\begin{center}
\begin{tabular}{r c c c c }
\hline\hline
\ & $a$ & $d$\\
\hline
\text{Constant wind}  & $2.45\times 10^{-4}\pm 7.6\times 10^{-5}$&$1.98\pm 0.03$ \\
\text{Variable wind}  & $9.68\times 10^{-4}\pm 1\times 10^{-4}$&$1.16\pm 0.09$ \\
\hline\hline
\end{tabular}
\end{center}
  
TABLE I. {\small {Parameters $a$ and $d$ of Eq.(\ref{fiteq}) fitted for both constant and variable winds.}}

\ 

An interesting observation is that if a second wind change occurs on the labyrinthic structure, $n_k$ is not affected. Once the diagonal structure is created, any come back to the transveral one is prohibited. The process is irreversible! However, if the initial topography is composed by ripples with a small amount of defects, the landscape can evolve to a nearly transversal structure. This behavior comes from the lack of kinks. Indeed, if $n_k$ is initialy small, a few number of labyrinths will be formed. As a consequence, the lanscape is less stable.

Moreover, the formation of labyrinthic structures induces a kind of ``memory effect''. Indee, asymptotic values $a$ listed in Table 1 are significantly different if one compares constant and variable cases! A difference in asymptotic values is a strong result supporting the idea that there is a memory of the wind direction on the landscape evolution. After a wind change, the evolution of the landscape depends essentialy on the former topography. Even after a long time, the surface always evolves in a way depending on its history, i.e on the number on wind orientation changes. The question is to know if real granular landscapes show this memory effect. This is left for future experimental work.

\section{Summary}

In summary, we have simulated unusual labyrinthic landscapes observed in earlier experiments. We have investigated the formation and evolution of these landscapes. We have demonstrated that the density of defects in a ripple structure is a relevant parameter to characterize the temporal evolution of such structures. Indeed, the number of ``defects'' present in the landscape decreases according to a negative power law of time. If wind orientation is changed, the power exponent shifts from a value $2$ to the value $1$. These exponents do not dependent on the occurence of wind change. We have also shown the emergence of a memory effect in the asymptotic value of the kink density. 

\vskip 1.0cm {\noindent \large Acknowledgements}

H. Caps is financially supported by the FRIA Belgium. This
work is also supported by the Belgian Royal Academy of Sciences
through the Ochs-Lefebvre prize.



\end{document}